\documentclass[journal]{IEEEtran}

\usepackage{cite}
\usepackage{amsmath,amssymb,amsfonts}
\usepackage{algorithmic}
\usepackage{algorithm}
\usepackage{graphicx}
\usepackage{textcomp}
\usepackage{xcolor}
\usepackage{url}
\usepackage{booktabs}
\usepackage{bm}

\begin{document}

\title{Belief-Aware Scheduling for Predictive Wildfire Hazard Mapping under Sparse-Window Telemetry}

\author{Xun~Shao,~\IEEEmembership{Senior~Member,~IEEE,}
        Kohsuke~Yamakawa,
        and~Cheah~Wai~Shiang%
\thanks{X. Shao and K. Yamakawa are with the Department of Electrical and Electronic Information Engineering, Toyohashi University of Technology, Toyohashi 441-8580, Japan (e-mail: shao.xun.ls@tut.jp; yamakawa.kosuke.wy@tut.jp).}%
\thanks{C. W. Shiang is with the Faculty of Computer Science and Information Technology, Universiti Malaysia Sarawak (UNIMAS), 94300 Kota Samarahan, Sarawak, Malaysia.}%
\thanks{Corresponding author: X. Shao (e-mail: shao.xun.ls@tut.jp).}}

\markboth{Shao \textit{et al.}: Belief-Aware Scheduling for Predictive Wildfire Hazard Mapping}{Shao \textit{et al.}: Belief-Aware Scheduling for Predictive Wildfire Hazard Mapping}

\maketitle

\begin{abstract}
An edge node monitoring a wildfire observes more than a duty-limited or windowed downlink can carry. The receiver must predict the $H$-step-ahead hazard map from whatever the link delivers. We argue the operative design problem is not which neural architecture to use but how to \emph{derive a structured belief sufficient for the receiver's prediction task} and \emph{maintain it through a scheduler that anticipates future transmission opportunities}. We formalise this as a partially observed sequential allocation problem with three coupled per-region action axes (sensing, representation, transmission), and derive each component of the structured belief from the $H$-step forward operator's input requirements. Identifying these mechanisms requires independent control over the window period $P$, per-window capacity $C$, predictive horizon $H$, and fuel composition, which is not separable in real-landscape data; we therefore evaluate on a physics-calibrated synthetic environment. Three empirical observations support the principle: the gap between a non-myopic activity-paced reference and uniform pacing is unimodal in window-period sparsity, peaking at intermediate spacing; ablating the structured belief, the dominant operative component flips between a default landscape (temporal staleness) and a structured landscape (static-risk prior), while the per-cell intensity belief is redundant in both; and a 40\,k-parameter lightweight cross-region attention encoder exceeds the FAIR activity-paced reference by $\sim$28\,\% on the default landscape and $\sim$11\,\% on the structured landscape. A deeper Transformer encoder does not improve over the lightweight encoder in mean predictive loss and exhibits higher training-seed variance. Within this task class and regime, a modest architectural inductive bias suffices when the belief and the scheduling problem are correctly posed.\end{abstract}


\section{Introduction}
\label{sec:introduction}
\IEEEPARstart{U}{nmanned} aerial platforms carrying radiometric thermal and RGB cameras, complemented by ground sensors for temperature, smoke, and wind, are increasingly used for early wildfire detection and tactical support. In the field, the return link to a relay or command post is frequently the binding constraint: long-range low-power radios such as LoRaWAN deliver kilobit-per-second throughput under strict duty-cycle regulation, and cellular or satellite uplinks (e.g., LEO constellations with windowed visibility) are intermittent in mountainous terrain, where network availability itself is a studied failure mode for disaster monitoring~\cite{brito2023data}. An edge node cannot forward every thermal frame. It must decide, under a tight communication budget that is sparse in time, where to sense, what to transmit, and at what fidelity, so that the command post can predict the evolving hazard map. This is the practical form of the shift from transmitting bits to transmitting task-relevant meaning advanced by goal-oriented and semantic communication~\cite{gunduz2023beyond}.

To make this concrete, consider a 4-hour daylight monitoring shift over an early-stage 16\,km $\times$ 16\,km wildfire incident ($256$\,km$^2$ area) discretised into a 16$\times$16 grid of 1\,km cells. The incident is partitioned into 25 actionable regions, each spanning roughly 9--10 cells. A single UAV can observe each region at a coarse or fine resolution, or skip it. The return link operates at LoRa-class throughput---about 28\,kB of airtime per 10\,min epoch under EU 868\,MHz at 1\,\% duty cycle---with windowed visibility from intermittent relay or LEO contacts. The command post predicts the hazard map $H$ epochs ahead. Three decisions must be made at every epoch: which regions to sense; for each region whether to send raw per-cell intensity, a per-cell summary, a per-region event flag, or nothing; and how to pace these decisions across windows whose budget is pooled but whose transmission opportunities are sparse in time. These are not three independent problems but a single allocation problem with three coupled axes, formalised below. Identifying the causal mechanisms this paper studies (Sections~\ref{sec:eval-tempcoup}--\ref{sec:eval-state}) requires independent control over $P$, $C$, $H$, and the fuel-type composition; these are not separable in real-landscape data, where fuel, terrain, wind, and ignition statistics co-vary. We therefore evaluate on a physics-calibrated synthetic environment (Section~\ref{sec:scenario}), with real-landscape evaluation as a distinct study (Section~\ref{sec:disc}).

\subsection{Design Principle}
\label{sec:principle}

The receiver's objective is a prediction: the H-step-ahead hazard map, scored against the realised fire intensity by mean-squared error. The scheduler does not, in general, need to transmit a high-fidelity reconstruction of the current fire to support this prediction. It needs to ensure that the receiver \emph{holds a belief sufficient for the H-step forward operator to predict accurately}, and to keep that belief sufficient under a sparse-window budget. The operative design problem can therefore be stated as a single principle, which is the thesis of this paper:

\begin{quote}
\textbf{Principle (Belief-Aware Scheduling for predictive telemetry under sparse-window constraints).} For a predictive task with horizon $H$ and forward operator $\Phi$, deployed on a single edge node with sparse, windowed transmission opportunities and a pooled budget across windows:
\begin{itemize}\itemsep0pt
\item[(i)] maintain a structured belief $B$ that is sufficient for $\Phi$ to predict at the required fidelity, with components \emph{derived from} $\Phi$'s input requirements and from the scheduler's allocation problem; and
\item[(ii)] maintain $B$ through a scheduling policy that anticipates future transmission opportunities.
\end{itemize}
For predictive wildfire hazard mapping in this regime, the derivation rule (Section~\ref{sec:belief}) yields the six-component tuple $B_t = (F, U, A^{\mathrm{obs}}, A^{\mathrm{tx}}, S, R)$. The framework of the derivation rule is task-class agnostic, but the claims in this paper---both the belief decomposition and the realisability findings---are validated only for this task class (single-drone predictive hazard mapping) and this regime (sparse-window pooled-budget telemetry); we do not assert that the same rule transfers without re-derivation to other predictive tasks or to different connectivity regimes.
\end{quote}

\begin{figure*}[!t]
\centerline{\includegraphics[width=0.92\textwidth]{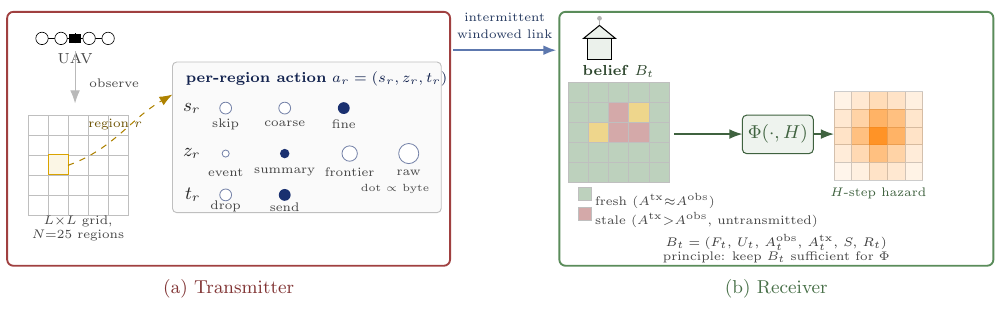}}
\caption{Scenario and scheduling principle (two panels, paper-specific differentiators only). \textbf{(a) Transmitter}: a single UAV observes an $L\!\times\!L$ grid partitioned into $N{=}25$ regions. For each region $r$ (callout) the on-board scheduler picks a graded \emph{three-axis} action $a_r = (s_r, z_r, t_r)$: sensing $s_r \in \{\text{skip, coarse, fine}\}$, representation $z_r \in \{\text{event, summary, frontier, raw}\}$ (dot size $\propto$ per-cell byte cost), and transmission $t_r \in \{\text{drop, send}\}$ (Section~\ref{sec:action}). Existing semantic-communication work typically chooses representation alone; the per-region three-axis structure is the differentiator of this work. \textbf{(b) Receiver}: data crosses a single intermittent windowed link to the command post, which maintains a structured belief $B_t = (F_t, U_t, A^{\mathrm{obs}}_t, A^{\mathrm{tx}}_t, S, R_t)$ (Section~\ref{sec:belief}). A region can be sensed at the drone yet not transmitted within the next window, so the drone-side observation age $A^{\mathrm{obs}}$ and the receiver-side transmission age $A^{\mathrm{tx}}$ diverge---visible in the belief map as a stale (red) cluster among fresh (green) regions, and indirectly encoding the consequence of sparse-window scheduling. The forward operator $\Phi(\cdot, H)$ maps the belief to an $H$-step-ahead hazard map; the principle is to keep $B_t$ sufficient for $\Phi$ at fidelity $L_H$, not to maximise reconstruction fidelity.}
\label{fig:scenario}
\end{figure*}

The principle does not depend on a particular neural architecture or semantic representation. The structured belief is also not an arbitrary feature list: each component is derived from a specific requirement of $\Phi$ or the scheduler. Within the scope just stated, the paper formulates the principle, identifies the empirical regime where it has bite, shows that the derived belief decomposition is active (the dominant operative components shift with landscape structure), and verifies that the principle is realisable with modest learned policy machinery. Fig.~\ref{fig:scenario} summarises the deployment scenario.

\subsection{Challenges of Sparse-Window Telemetry}
\label{sec:why-sparse}

Per-epoch greedy heuristics---natural under linear per-epoch resource constraints---can be sub-optimal when transmission opportunities are sparse in time (windowed visibility) and resources are pooled across the episode rather than per-epoch: a budget spent in an early window can no longer be recovered, and preserving budget for high-activity future windows requires anticipating temporal coupling across the $K$-epoch horizon. A basic question is therefore for what range of window-period sparsity this temporal coupling creates structure that a non-myopic scheduler can exploit. Section~\ref{sec:eval-tempcoup} characterises this regime by sweeping the window period $P$; the principle of Section~\ref{sec:principle} is expected to have empirical room only where the activity-paced--uniform gap is non-negligible.

\subsection{Belief Components for the Predictive Task}
\label{sec:which-belief}

Given the structured belief tuple $B_t = (F, U, A^{\mathrm{obs}}, A^{\mathrm{tx}}, S, R)$ derived from the requirements of $\Phi$, a natural question is what information the scheduler actually relies on, and whether this depends on the landscape regime in which the scheduler operates. Section~\ref{sec:eval-state} addresses this through a per-component ablation on two landscapes (a default landscape, and a structured landscape with deterministic fuel belts and ridges); a derived decomposition would be evidenced by an operative shift across regimes rather than a single dominant feature.

\subsection{Realising the Principle with Learned Policies}
\label{sec:arch-not-thesis}

The principle of Section~\ref{sec:principle} is architecture-agnostic; a separate question is what level of policy expressiveness is required to realise it. Section~\ref{sec:eval-arch} compares four learned policy architectures (a memoryless state-rich MLP, a GRU, a lightweight cross-region attention encoder, and a deeper Transformer) under identical online PPO training on the default and structured landscapes. The role of this comparison is a realisability check rather than a claim that any particular inductive bias is universally required.

\subsection{Evaluation Considerations}
\label{sec:fair-unfair-intro}

A methodological concern in evaluating belief-only learned policies is the choice of reference baseline: a baseline that allocates against the privileged true fire field is unrealisable at deployment, and using it as the reference can inflate the apparent learning improvement. Section~\ref{sec:eval-fair} quantifies this distinction (FAIR vs.\ UNFAIR baselines); throughout this paper we report learned-policy results against a deployable belief-only baseline.

\subsection{Contributions}

The paper makes the following contributions. (i) We \emph{formulate} belief-aware scheduling as a derivation-rule principle: given a forward operator $\Phi$, the structured belief is obtained as a derivation from $\Phi$'s input requirements and the scheduler's allocation constraints, rather than as an engineered feature list (Sec.~\ref{sec:principle},~\ref{sec:form}). (ii) We \emph{characterise} the empirical regime in which the principle has bite, by sweeping the window period $P$ and identifying where the temporal-opportunity coupling exposes learnable structure (Sec.~\ref{sec:eval-tempcoup}). (iii) We \emph{establish} the empirical activity of the derived belief decomposition through a per-component ablation on two landscapes (Sec.~\ref{sec:eval-state}). (iv) We \emph{verify} realisability via a controlled multi-seed comparison of four learned policy architectures under identical online PPO training, framed as a robustness check on the principle (Sec.~\ref{sec:eval-arch}). (v) We \emph{quantify} the inflation produced by unfair (privileged-truth) baselines as a methodological note for fair-comparison reporting (Sec.~\ref{sec:eval-fair}).

The remainder of this paper is organised as follows. Section~\ref{sec:related} situates the work. Section~\ref{sec:form} formalises the structured belief and action axes. Section~\ref{sec:runtime} derives the device-grounded budgets and Section~\ref{sec:policies} defines the baseline and learned policies. Section~\ref{sec:eval} reports the empirical evidence in the order of the principle (regime $\to$ mechanism $\to$ realisability $\to$ honest comparison). Section~\ref{sec:disc} discusses limitations and Section~\ref{sec:conc} concludes.

\section{Related Work}
\label{sec:related}

\subsection{Goal-Oriented and Semantic Communication}
A growing body of work transmits task-relevant content rather than reconstructed signals, allocating rate to information that affects a downstream objective~\cite{gunduz2023beyond}. Context-aware semantic communication selects high-impact content under bandwidth constraints using a learned gating mechanism over multiple data modalities~\cite{liu2025context}; task-adaptive schemes transmit a coarse representation and then regenerate only decision-relevant detail at the receiver under task feedback~\cite{guo2025task}; and the rate--distortion--classification view characterises representations that simultaneously preserve reconstruction and a downstream decision under a rate budget~\cite{nguyen2025universal}. These works treat representation choice as the central design lever, for a single transmitter--receiver pair. Our problem differs in three structural ways: the transmitter additionally decides \emph{where to sense}; it must spend one budget \emph{across space and time}, so that improving one region's telemetry forgoes another's, possibly across future windows; and the representation is scored by a spatial-temporal predictive task. The design lever in our framing is not the representation alone but the joint compression-and-scheduling that keeps the receiver's belief sufficient for prediction.

\subsection{POMDP Active Perception and Policy Architectures}
Decision-theoretic active sensing surveys note that ordinary Markov rewards cannot credit an agent for knowing more~\cite{veiga2023reactive}. The POPGym benchmark suite~\cite{morad2023popgym} establishes recurrent (GRU/LSTM/Transformer) hidden state as the standard belief representation in online POMDP RL, with training budgets in the millions of environment steps. Information-gain shaping rewards~\cite{hu2024maxinforl} and asymmetric training with privileged information~\cite{wang2024active} extend active perception to high-dimensional control. We adopt a complementary stance: rather than treating recurrent hidden state as the only belief substrate, we provide an explicit structured belief tuple (hazard estimate, uncertainty, observation/transmission age, static prior, resource state) and a per-component ablation. The contribution is the derivation rule and the ablation, not a head-to-head performance comparison against implicit-belief baselines. Decision Transformer~\cite{chen2021decisiontransformer} and related return-conditioned offline methods address offline sequential decision making and are out of scope.

\subsection{Edge AI for Wildfire Monitoring}
On-device detection for UAV- and sensor-based wildfire monitoring is active and largely deployed. Recent IEEE Access work spans real-time forest-fire detection with explainable CNN heads on UAV imagery~\cite{alam2025firenet}, integrated drone systems pairing detection with downstream risk prediction~\cite{lelis2024drone}, and a survey of AI-enabled IoT pipelines that covers prediction, detection, and post-operations~\cite{badary2026survey}. Related work runs autonomous surveillance with machine vision plus control~\cite{bonilla2025forest}, onboard segmentation distilled from larger teachers~\cite{pesonen2024detecting}, and transfer-learned detectors at the edge~\cite{vazquez2025detecting}. Resource-aware UAV--edge frameworks co-optimise route planning, fleet sizing, and edge placement~\cite{huang2026risk}, and adaptive-edge formulations point out that deployed inference reconfigures to time-varying budgets~\cite{pittorino2026adaptive}. Most of this work runs detection and transmits alerts. The question we study is different: a graded per-region sensing--representation--transmission allocation scored by a receiver-side predictive task under a shared windowed budget.

\subsection{Wildfire Prediction Benchmarks}
WildfireSpreadTS~\cite{gerard2023wsts} (607 fire events, 13,607 daily images at VIIRS 375\,m resolution), Mesogeos~\cite{kondylatos2025mesogeos} (a 17-year Mediterranean dataset at 1\,km resolution), and the Next-Day Wildfire Spread benchmark~\cite{huot2022nextday} provide large-scale benchmarks for binary burn-probability prediction from real satellite data. We focus on the orthogonal problem of belief-aware $S\times R\times T$ scheduling for a single-drone active sensor, evaluated on physics-calibrated synthetic fires; our primary metric is continuous-intensity MSE, with secondary binary metrics (F1/IoU/Dice) reported for benchmark comparability.

\subsection{Reinforcement Learning for Resource Allocation}
Deep reinforcement learning is widely applied to resource allocation, scheduling, and access control in communications and networking~\cite{luong2019applications}, including LoRaWAN spreading-factor and transmit-power allocation~\cite{wang2025online}. A recurring concern is interpretability~\cite{rusu2016policy}. Our work contributes a controlled architecture ablation under identical PPO training. The architecture comparison is reported as a realisability check on a design principle, not as evidence that a particular inductive bias is universally necessary.

\section{Problem Formulation}
\label{sec:form}

This section formalises the design principle. Concrete experimental values are collected in Section~\ref{sec:eval}; the notation is summarised in Table~\ref{tab:notation} and the percentage definitions used throughout in Table~\ref{tab:percentdef}.

\begin{table*}
\caption{\textbf{Notation.}}
\label{tab:notation}
\setlength{\tabcolsep}{6pt}
\centering
\footnotesize
\begin{tabular}{|l|p{2.7in}|l|p{2.6in}|}
\hline
Symbol & Meaning & Symbol & Meaning\\
\hline
$L$ & grid linear dimension & $B_w$ & per-window airtime budget (bytes)\\
$N$ & number of coarse regions & $E_w$ & per-window flight-energy budget (J)\\
$K$ & number of decision epochs & $H$ & predictive horizon (epochs)\\
$P$ & window period (epochs between windows) & $\Phi(\cdot, H)$ & forward operator (\,$H$-step hazard predictor)\\
$C$ & per-window capacity multiplier & $F_t$ & hazard-estimate field (belief, $L\!\times\!L$)\\
$\mathcal{C},\mathcal{C}_r$ & cells; cells of region $r$ & $U_t$ & belief-uncertainty proxy field\\
$s_r,z_r,t_r$ & sensing, representation, transmission for region $r$ & $A^{\mathrm{obs}}_t, A^{\mathrm{tx}}_t$ & per-region observation/transmission age\\
$a_r$ & per-region action $(s_r,z_r,t_r)$ & $S_r$ & per-region static-risk prior\\
$\Pi$ & feasibility projection & $R_t$ & resource-state vector (4-dim)\\
$L_H(t)$ & predictive MSE at horizon $H$ & $B_t$ & structured belief tuple\\
F1, IoU, Dice & secondary binary metrics & $\theta_{\mathrm{burn}}$ & binary-metric threshold\\
\hline
\end{tabular}
\end{table*}

\begin{table*}
\caption{\textbf{Percentage definitions} used in tables, figures, and prose. For policy $A$ with mean predictive loss $L_A$, evaluated against FAIR uniform $L_u$ and the FAIR activity-paced reference $L_o$. $L_o$ combines belief-only scoring with privileged per-window pacing access; it is a comparison anchor, not a performance upper bound, so gap closure $>100\,\%$ is expected when a learned policy infers a better pacing rule from the belief.}
\label{tab:percentdef}
\setlength{\tabcolsep}{6pt}
\centering
\footnotesize
\begin{tabular}{|l|l|p{3.3in}|}
\hline
Name & Definition & First used / typical magnitude\\
\hline
Relative gap & $(L_u - L_o)/L_u$ & Table~\ref{tab:tempcoup}; e.g.\ 54\,\% at $P=2$\\
Gap closure & $(L_u - L_A)/(L_u - L_o)$ & Tables~\ref{tab:arch},~\ref{tab:arch_str}; e.g.\ 144.5\,\% (Lightw.\,Attn., default)\\
Vs.\ FAIR uniform & $(L_u - L_A)/L_u$ & Table~\ref{tab:arch}; e.g.\ $+56$\,\% (Lightw.\,Attn., default)\\
Excess over activity-paced ref. & $(L_o - L_A)/L_o$ & Abstract / Sec.~\ref{sec:arch-not-thesis}; e.g.\ $\sim$28\,\% (Lightw.\,Attn., default)\\
Inflation factor & $L_{\mathrm{FAIR,unif}} / L_{\mathrm{UNFAIR,unif}}$ & Sec.~\ref{sec:eval-fair}; e.g.\ $2.5\times$\\
\hline
\end{tabular}
\end{table*}

\subsection{Scenario and Physics-Calibrated Synthetic Environment}
\label{sec:scenario}

The causal mechanisms studied in this paper---the temporal-opportunity coupling (Section~\ref{sec:eval-tempcoup}) and the operative shift of belief components across landscape regimes (Section~\ref{sec:eval-state})---require independent control over the window period $P$, the per-window capacity $C$, the predictive horizon $H$, and the fuel-type composition. These four cannot be varied independently in real-landscape data, where fuel, terrain, wind, and ignition statistics co-vary by construction. A physics-calibrated synthetic environment supports this independent control while keeping the dynamics anchored in operational fire-behaviour models (Rothermel rate-of-spread; NWCG-calibrated fuel-model rates; Anderson-13 / FBFM40 fuel-type prevalence), at the $1$\,km cell resolution used by the Mesogeos benchmark~\cite{kondylatos2025mesogeos}.

We consider a single wildfire incident over an area discretised into an $L \times L$ grid of square cells, partitioned into $N$ coarse regions on a regular sub-grid. The fire propagates by a Rothermel-calibrated continuous-intensity model. Per-cell intensity $f_t(c) \in [0,1]$ and residual fuel $\phi_t(c) \in [0,1]$ evolve at sub-minute timesteps:
\begin{equation}
R(c) = R_0(v_c)\,\eta_M\,(1 + 0.7\,\|\mathbf{w}\| + 5.275\,\|\mathbf{g}\|^2),
\label{eq:ros}
\end{equation}
\begin{equation}
\Psi(c) = \!\!\sum_{c' \in \mathcal{N}(c)}\!\! \omega(\mathbf{r}_{c c'}, \mathbf{w}, \mathbf{g})\, \frac{f_t(c')\,R(c')}{\ell},
\label{eq:flux}
\end{equation}
\begin{equation}
\Pr[c\text{ ignites}] = (1 - \mathbb{1}[f_t(c) \ge \epsilon_b])(1 - \exp(-\Psi(c)\Delta t)),
\label{eq:ignite}
\end{equation}
where $R_0(\cdot)$ is the base rate of spread per fuel model (NWCG-calibrated: NB/GR/SH/TL/TU at $0/6/3/1.5/2.5$\,m/min, anchored to NWCG operational standards~\cite{nwcg_ros}); $\eta_M$ is the moisture-suppression factor; $\mathbf{w}$ and $\mathbf{g}$ are the wind and slope vectors; $\omega(\cdot)$ is the directional Rothermel weight; $\ell$ is the cell side; $\epsilon_b = 0.05$ is the \emph{ignition threshold} of the physics model (distinct from the binary-metric threshold $\theta_{\mathrm{burn}} = 0.2$ used in Section~\ref{sec:objective} for the secondary F1/IoU metrics); and $\Delta t = 1$\,min is the simulation substep. The landscape is a physics-calibrated synthetic environment generated with Anderson-13 / FBFM40-inspired fuel-type prevalence (NB 8\,\%, GR 22\,\%, SH 28\,\%, TL 28\,\%, TU 14\,\%). The decision epoch length $\Delta$, the horizon $K$, and ignition seeds NIGN are experimental choices and are listed with the other settings in Section~\ref{sec:eval-settings}.

\subsection{Structured Belief as Derivation from $\Phi$'s Requirements}
\label{sec:belief}

The hidden state at epoch $t$ is $(\mathbf{f}_t, \boldsymbol{\phi}_t)$. The receiver does not observe this directly; it maintains a structured belief
\begin{equation}
B_t = (F_t,\; U_t,\; A^{\mathrm{obs}}_t,\; A^{\mathrm{tx}}_t,\; S,\; R_t).
\label{eq:belief}
\end{equation}
We do not present $B_t$ as a hand-picked feature list. Each component is derived from a specific requirement of either the H-step forward operator $\Phi(\cdot, H)$ (which must predict accurately) or the scheduler (which must allocate forward in time under sparse-window constraints):
\begin{itemize}
\item $F_t \in [0,1]^{L \times L}$ --- \emph{per-cell hazard estimate.} Required input to $\Phi$ for any per-cell prediction. Updated by belief filtering between transmissions and by reconstruction when a region is transmitted.
\item $U_t = |F_t - F_{t-1}|$ --- \emph{predictive-uncertainty proxy.} Quantifies how much the receiver's belief has changed; informs whether $\Phi$'s output is stable or drifting. Required to assess prediction confidence.
\item $A^{\mathrm{obs}}_t \in \mathbb{N}^N$ --- \emph{per-region observation age.} The drone-side belief-update freshness. Prediction error grows monotonically with this quantity under any non-trivial belief filter; required so the scheduler can target sensing at regions whose belief has staled.
\item $A^{\mathrm{tx}}_t \in \mathbb{N}^N$ --- \emph{per-region transmission age.} Distinct from $A^{\mathrm{obs}}_t$: a region can be sensed at the drone but un-transmitted to the receiver. Required so the scheduler can target transmission at regions whose receiver-side belief is stale even when its drone-side belief is fresh.
\item $S \in [0,1]^N$ --- \emph{static-risk prior.} Encodes landscape-level susceptibility (fuel-type prevalence, terrain). Required so the scheduler can anticipate fires not yet observed in any region.
\item $R_t \in [0,1]^4$ --- \emph{resource state} (remaining airtime/energy, time to next window, remaining windows). Required so the scheduler can allocate budget forward in time over the sparse-window horizon.
\end{itemize}
$B_t$ is a practical sufficient representation for a scheduler whose objective is $L_H$: each component is motivated by a specific input requirement of $\Phi$ or the scheduler's allocation problem, and the per-component ablation in Section~\ref{sec:eval-state} verifies empirically that removing any component degrades scheduler performance. Section~\ref{sec:eval-state} reports a per-component ablation. Each component is kept as a point estimate rather than a full posterior distribution; extension to Bayesian belief filtering is a natural follow-up that refines the realisation rather than the derivation rule.

\subsection{Action Axes: Sensing, Representation, Transmission}
\label{sec:action}

The per-epoch action is $a_t = \{(s_r, z_r, t_r)\}_{r=1}^N$, with three axes per region:
\begin{align}
s_r &\in \mathcal{S} = \{\mathrm{skip},\,\mathrm{coarse},\,\mathrm{fine}\}, \nonumber\\
z_r &\in \mathcal{Z} = \{\mathrm{event},\,\mathrm{summary},\,\mathrm{frontier},\,\mathrm{raw}\},\label{eq:action}\\
t_r &\in \mathcal{T} = \{\mathrm{drop},\,\mathrm{send}\}. \nonumber
\end{align}
\emph{Sensing} $s_r$ determines the resolution at which the drone observes region $r$. \emph{Representation} $z_r$ determines the encoding of the sensed signal: event (per-block binary flag, 0.125 B/cell), summary (block-mean quantized, 0.25 B/cell), frontier (2.0 B/cell), or raw (per-cell intensity, 4 B/cell). \emph{Transmission} $t_r$ determines whether the region is sent in this epoch's window or held. The per-cell byte costs follow an information-theoretic 1\,:\,2\,:\,16\,:\,32 ratio (event:summary:frontier:raw).

\subsection{Predictive Hazard Mapping Objective}
\label{sec:objective}

The receiver consumes the (possibly partial) reconstruction at epoch $t$ to maintain $F_t$ and predicts the fire intensity field $H$ epochs ahead via a forward operator $\Phi(\cdot, H)$ that applies the Rothermel dynamics for $H$ epochs. The primary per-epoch loss is the mean squared error of the prediction against the realised future:
\begin{equation}
L_H(t) = \frac{1}{|\mathcal{C}|}\big\|\Phi(F_t, H) - \mathbf{f}_{t+H}\big\|_2^2,
\label{eq:lh}
\end{equation}
and the episodic objective is
\begin{equation}
\min_{\pi}\;\; \mathbb{E}\!\left[\sum_{t=0}^{K-1} L_H(t)\right].
\label{eq:obj}
\end{equation}
Secondary per-epoch binary metrics on the thresholded hazard map---F1, IoU, Dice at $\theta_{\mathrm{burn}} = 0.2$---are reported as a robustness check and for benchmark comparability with binary-classification wildfire benchmarks.

\section{Device-Grounded Budgets and Sparse-Window Transmission}
\label{sec:runtime}

Two shared resource budgets bind every per-epoch action, both derived from device specifications. The full episode has aggregate budgets $B_{\mathrm{ep}}^{\mathrm{TX}}$ (airtime) and $E_{\mathrm{ep}}^{\mathrm{TX}}$ (energy), pooled across $K$ epochs and divided across $W = \lceil K/P \rceil$ transmission windows.

\textbf{Airtime budget.} The duty-limited time-on-air follows the LoRa time-on-air model for the EU 868\,MHz band~\cite{loraalliance2022rp,semtech_an120013}. At 1\,\% duty cycle and the highest LoRa configuration (37.5\,kbps),
\begin{equation}
B_{\mathrm{epoch}} = \frac{R\,\Delta\,\eta}{8} \approx 28{,}125 \text{ B/epoch.}
\label{eq:lora}
\end{equation}
The effective per-epoch budget used in this paper, with the per-window capacity multiplier $C$ defined below, is reported with the other experimental settings in Section~\ref{sec:eval-settings}.

\textbf{Energy budget.} Active sensing consumes flight power: with battery capacity $W_{\mathrm{bat}} = 131.6$\,Wh (DJI M30T TB30 pair) and endurance $T_{\mathrm{end}} = 41$\,min, the flight power is $P_{\mathrm{fly}} \approx 192.6$\,W, giving a per-epoch envelope $E_{\mathrm{epoch}} \approx 115.6$\,kJ at $\Delta = 10$\,min. Per-region sensing dwell consumes $\tau_{\mathrm{coarse}} = 30$\,s and $\tau_{\mathrm{fine}} = 90$\,s. In the experimental regime we evaluate ($B_w/P \approx 200$\,B/epoch, see Section~\ref{sec:eval-settings}), the airtime budget binds: full-region transmission saturates $B_w$ before sensing dwell saturates $E_w$. The energy budget is derived to keep the device-side accounting honest and to ensure feasibility under per-window dwell limits, but it is not the binding constraint at the operating point reported below.

\textbf{Sparse-window transmission: a hierarchy of models.} Intermittent transmission opportunity can be modelled at four levels of increasing realism: (i) full DTN contact plans with scheduled satellite/relay contacts, operator-specific but deployment-realistic for LEO and store-and-forward systems; (ii) trace-driven availability from measured LoRa/LTE/LEO visibility logs, empirically grounded but generalisation-limited; (iii) stochastic availability with $A_t \in \{0,1\}$ and per-window capacity $B_t \ge 0$ sampled from a (semi-)Markov chain, theoretically tractable and common in AoI / task-oriented satellite work; and (iv) periodic windows ($t \equiv 0 \pmod P$, capacity $B_w = C \cdot P \cdot B_{\mathrm{epoch}}$), a deterministic abstraction used for controlled mechanism isolation. In this paper we use (iv) as the primary setting so that the opportunity-coupling mechanism (Section~\ref{sec:eval-tempcoup}) can be isolated from confounding stochastic variance; the specific $P$ and $C$ sweeps are reported in Section~\ref{sec:eval-settings}. Sparse-window transmission is precisely what makes opportunity-aware scheduling non-trivial: budget left unused in a window is forfeited; budget committed early cannot be recovered when a future window's activity is high. Tiers (i)--(iii) condition deployment claims and are future work; in particular, a stochastic-availability robustness check (tier iii) with $E[\text{windows}]$ matched to the periodic setting is the most natural follow-up.

\textbf{Feasibility projection.} Every policy passes through a feasibility projection $\Pi$ that respects (a) per-drone energy, (b) per-window airtime, and (c) the windowed-transmission constraint. $\Pi$ is applied identically to every policy, so no policy can exceed budgets and feasibility cannot account for any performance difference.

\section{Policies}
\label{sec:policies}

\subsection{Non-Learned Baselines: FAIR vs.~UNFAIR}
\label{sec:baselines}

Whether a baseline uses information the learned policy cannot use is a methodological pitfall in this literature. We define two classes.

\textbf{UNFAIR (privileged truth).} The classical greedy scheduler that selects the region--action with the largest immediate gain by comparing the candidate reconstruction against the \emph{true} fire field $\mathbf{f}_t$. Not a deployable baseline.

\textbf{FAIR (belief-only).} The same greedy scheduler, but scoring each region by a belief-derived priority that combines $U_t$, the F-max in the region, $A^{\mathrm{obs}}_t$, and $S$. Uses only quantities a deployed policy can observe; the appropriate comparison for a learned scheduler.

For each class we instantiate three pacing variants: \emph{uniform} (budget split equally across windows), \emph{eager} (all budget at the first window), and \emph{activity-paced} (per-window budget proportional to cumulative ground-truth activity in the upcoming inter-window interval). The central reference for learned policies is FAIR uniform; the \emph{FAIR activity-paced reference} combines belief-only scoring with privileged per-window pacing access and is therefore a non-trivial comparison anchor, not a performance upper bound: a learned policy that infers activity from the belief alone can exceed it, and gap closure above 100\,\% in subsequent tables is the expected signature of such improvement.

\subsection{Learned Policy Architectures (Four Realisations of the Principle)}
\label{sec:archs}

We train four policy architectures under identical online PPO. They are not competing principles but four realisations of the belief-aware scheduling principle of Section~\ref{sec:principle}, differing only in how the structured belief is mixed into per-region action heads.

\textbf{MLP} ($\sim$167\,k params). Flattened $B_t$ feeds two 200-unit hidden layers and per-region action heads. The simplest realisation.

\textbf{GRU} ($\sim$90\,k params). 80-unit GRU cell over the per-epoch flattened state. Tests temporal recurrence on top of $B_t$.

\textbf{Lightweight Attention} ($\sim$40\,k params). 2-layer transformer encoder, 4 heads, embedding dim 48, treating per-region belief components as a token sequence plus a global token.

\textbf{Transformer} ($\sim$202\,k params). 4 layers, $d = 64$, 4 heads. Tests whether depth on top of cross-region mixing helps.

All architectures share input representation, action factorisation, training, and feasibility projection. Only the policy network varies.

\subsection{Training: BC Warm-Start + PPO}
\label{sec:training}

For each architecture: (i) warm-start by behaviour cloning on demonstrations from the FAIR-uniform baseline for $N_{\mathrm{BC}}$ episodes; (ii) fine-tune by PPO with GAE for $N_{\mathrm{PPO}}$ episodes on $r_t = -L_H(t)$. Clipped-surrogate PPO with $\gamma = 0.99$, $\lambda = 0.95$, clip $\varepsilon = 0.2$, entropy bonus 0.01, learning rate $3 \times 10^{-4}$, 4 optimisation epochs per update over batches of 4 episodes. We evaluate at multiple budgets to expose the budget-dependence of the architecture ranking.

\section{Evaluation}
\label{sec:eval}

The evaluation is organised in the order of the principle: first the regime (Section~\ref{sec:eval-tempcoup}) in which the principle has empirical room; then the mechanism (Section~\ref{sec:eval-state}) that exposes which belief components actually drive the scheduler; then the realisability check (Section~\ref{sec:eval-arch}) that the principle can be implemented at modest cost; then the honest-comparison observation (Section~\ref{sec:eval-fair}); and finally the budget dependence (Section~\ref{sec:eval-budget}) and horizon ablation (Section~\ref{sec:eval-horizon}).

\subsection{Experimental Settings}
\label{sec:eval-settings}

\textbf{Spatial setting.} We instantiate the formulation of Section~\ref{sec:form} on a $16 \times 16$ grid of $1$\,km cells, giving a $16$\,km $\times 16$\,km area ($256$\,km$^2$) at a spatial resolution matching the Mesogeos benchmark~\cite{kondylatos2025mesogeos}. The grid is partitioned into $N = 25$ coarse regions on a $5 \times 5$ arrangement, with $\sim$10 cells per region.

\textbf{Temporal setting.} Each decision epoch is $\Delta = 10$\,min long; we use $K = 25$ epochs (a 4-hour daylight monitoring shift) and ignition seeds NIGN $\in \{2, 3\}$.

\textbf{Connectivity setting.} The sparse-window connectivity (Section~\ref{sec:runtime}) is swept over the window period $P \in \{1, 2, 4, 8\}$ and the per-window capacity multiplier $C \in \{0.5, 1.0, 2.0\}$. To model a contention-heavy multi-UAV scenario, we deliberately bind the per-epoch airtime well below the LoRa-max derivation of Section~\ref{sec:runtime}: with 25 regions of $\sim$10 cells each and an information-theoretic 1:2:16:32 ratio (event:summary:frontier:raw, B/cell), the binding regime sits at $B_w / P \approx 200$\,B/epoch (roughly $1/140$ of the $28{,}125$\,B/epoch LoRa-max budget). At this level, raw all-region transmission is marginally infeasible and the representation choice becomes operative.

\textbf{Seeds.} Training and evaluation use disjoint seed pools (training $\in [200, 500)$; test $\in [100, 140)$), with $40$ paired test seeds per architecture per setting. Each trained policy is evaluated twice on the same $40$ test seeds---once deterministically (argmax over the policy's action distribution) and once stochastically (sampled actions)---and the reported $L_H$ is the lower of the two mean values, denoted ``best of det/stoch''. The two evaluation modes are reported because the optimal exploration--exploitation balance at evaluation time differs by architecture and by PPO entropy setting; taking the lower of det and stoch absorbs this incidental difference and avoids penalising an architecture whose policy distribution happens to be sharper or flatter than another's. Differences are reported with standard errors; bold marks effects of $5$\,SE or more.

\textbf{Fixed-action arm (d*).} For the temporal-opportunity sweep in Section~\ref{sec:eval-tempcoup}, we evaluate a fixed-action variant labelled \emph{d*}: every region is sensed at fine resolution ($s = \mathrm{fine}$), encoded as a quantized block-mean summary ($z = \mathrm{summary}$), and always transmitted ($t = \mathrm{send}$). This holds the per-region $(s, z, t)$ choice constant across regions, so that only the per-window budget allocation varies; the FAIR uniform--activity-paced gap on the d* arm therefore isolates the pure pacing-mechanism contribution and is the cleanest probe of the temporal-opportunity coupling.

\subsection{Regime: Temporal Opportunity Coupling is Unimodal in $P$}
\label{sec:eval-tempcoup}

The principle's empirical room is the gap a non-myopic, opportunity-aware scheduler has over a uniform-pacing baseline. We sweep window period $P \in \{1,2,4,8\}$ at fixed $C = 1.0$ and $H = 3$, comparing FAIR uniform against FAIR activity-paced reference.

\begin{table}
\caption{\textbf{The principle's empirical room: FAIR activity-paced reference--uniform gap as a function of window period $P$ on the d* arm (fine sensing + summary representation), at $C=1.0$, $H=3$, $K=25$, 40 paired test seeds. The relative gap is $\mathrm{gap}/L_u$ ($=$ relative improvement of activity-paced over uniform pacing); the gap is non-monotonic in $P$ and peaks at intermediate $P=2$.}}
\label{tab:tempcoup}
\setlength{\tabcolsep}{4pt}
\centering
\footnotesize
\begin{tabular}{|c|r|r|r|r|}
\hline
$P$ & FAIR uniform $L_H$ & FAIR-AP ref.\ $L_H$ & gap & rel.\,gap\\
\hline
1 & 0.111\,$\pm$\,0.009 & 0.060\,$\pm$\,0.004 & 0.052 & 46\,\%\\
\textbf{2} & \textbf{0.135}\,$\pm$\,0.011 & \textbf{0.063}\,$\pm$\,0.005 & \textbf{0.073} & \textbf{54\,\%}\\
4 & 0.156\,$\pm$\,0.012 & 0.086\,$\pm$\,0.006 & 0.069 & 44\,\%\\
8 & 0.162\,$\pm$\,0.013 & 0.145\,$\pm$\,0.009 & 0.016 & 10\,\%\\
\hline
\end{tabular}
\end{table}

\begin{figure}[!t]
\centering
\includegraphics[width=\columnwidth]{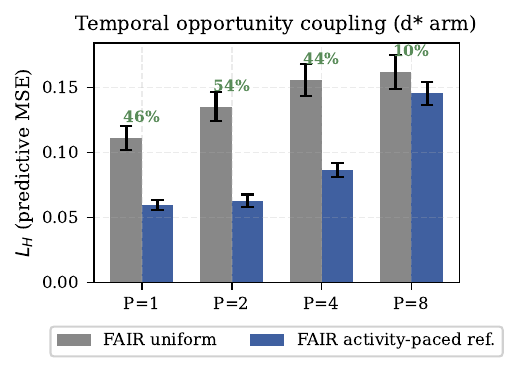}
\caption{\textbf{Empirical room for the principle. The FAIR activity-paced reference--uniform gap is non-monotonic in window period $P$ on the d* arm: 46\,\% at $P=1$, peaking at 54\,\% at $P=2$, 44\,\% at $P=4$, and dropping to 10\,\% at $P=8$. Both bars at each $P$ use the same FAIR (belief-only) logic; only the per-window budget allocation differs. The relative gap is $\mathrm{gap}/L_u$ ($=$ improvement of activity-paced over uniform pacing).}\label{fig:tempcoup}}
\end{figure}

\textbf{Result (Table~\ref{tab:tempcoup}, Fig.~\ref{fig:tempcoup}).} The FAIR activity-paced reference--uniform gap is non-monotonic in $P$: 46\,\% at $P = 1$, peaks at $P = 2$ (54\,\% relative improvement, 10.7\,SE on the paired difference; windows sparse enough to require pacing but dense enough to allow flexible reallocation), 44\,\% at $P = 4$, and shrinks to 10\,\% at $P = 8$ (both activity-paced and uniform are equally constrained by sparsity). The peak at intermediate $P$ identifies the deployment regime in which opportunity-aware scheduling has the most room, and the rest of the evaluation operates inside this regime ($P = 2$, $C = 1.0$, $H = 3$).

\textbf{Note on baseline values across action regimes.} The FAIR uniform $L_u$ and FAIR activity-paced reference $L_o$ values differ between the d* arm (Table~\ref{tab:tempcoup}, $L_u=0.135$, $L_o=0.063$ at $P=2$) and the full per-region action space (Tables~\ref{tab:arch} and~\ref{tab:fair}, $L_u=0.150$, $L_o=0.093$ at the same operating point). The d* arm fixes the per-region representation to summary, which preserves belief quality, whereas the full action space admits skip/coarse and event/frontier/raw choices under the same belief and pacing rule, raising $L_u$ and shifting $L_o$ correspondingly. The two regimes are not in conflict; they probe different action sets. All learned-policy comparisons (Tables~\ref{tab:arch},~\ref{tab:arch_str},~\ref{tab:budget}) and the FAIR/UNFAIR comparison (Table~\ref{tab:fair}) use the full action space.

\subsection{Mechanism: Which Belief Components Drive the Scheduler}
\label{sec:eval-state}

We test the first half of the principle (maintaining a structured belief sufficient for $\Phi$ to predict) by asking which components of $B_t$ actually carry the action-relevant signal, and whether the answer itself depends on landscape structure. The MLP is retrained with one component removed at a time, holding the training budget (1200 episodes) and the operating point ($P=2$, $C=1.0$, $H=3$) fixed. The MLP is used because, without cross-region attention or recurrent memory, its performance is most directly a function of the input state. The ablation runs on two landscape generators: a \emph{default (smoothed-noise) landscape}, the setting of Sections~\ref{sec:eval-tempcoup}--\ref{sec:eval-arch}, and a \emph{structured-synthetic landscape} with deterministic fuel belts and a ridge corridor (defined below). We refer to these as ``default'' and ``structured'' throughout.

\begin{figure}[!t]
\centering
\includegraphics[width=\columnwidth]{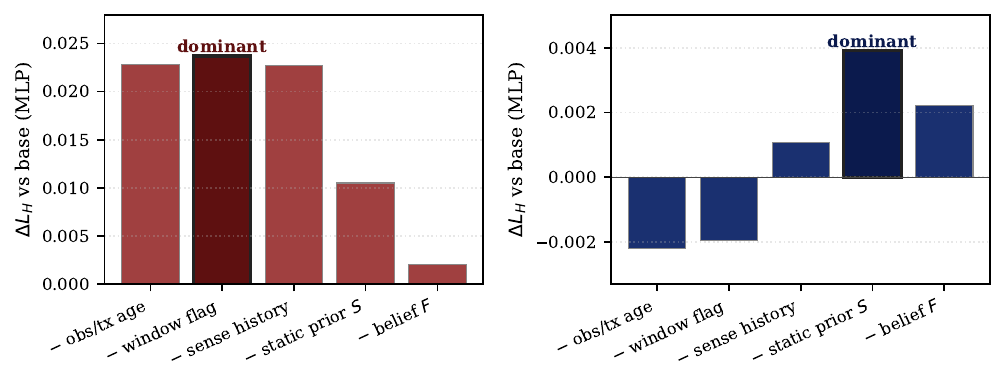}
\caption{\textbf{Mechanism. Operative belief components flip with landscape structure. \textit{(a) Left, default landscape}: removing the temporal-staleness components is the most costly ($+0.023$, dominant component highlighted). \textit{(b) Right, structured landscape (zoomed y-axis)}: the static-risk prior $S$ becomes dominant ($+0.004$); the temporal-staleness components are uninformative or mildly harmful. The per-cell hazard estimate $F_t$ remains redundant in both regimes. This is the empirical signature that $B_t$'s decomposition tracks problem structure, not arbitrary feature engineering.}\label{fig:state_flip}}
\end{figure}

\textbf{Result on default landscape (Table~\ref{tab:state}, red bars in Fig.~\ref{fig:state_flip}).} The MLP base reaches $L_H = 0.110$ (70\,\% gap closed). Removing the observation and transmission age maps ($A^{\mathrm{obs}}_t, A^{\mathrm{tx}}_t$) is the most costly ($+0.023$, gap closure drops to 31\,\%); removing the in-window flag ($+0.024$) and the per-region sense-type history ($+0.023$) are comparable. The static-risk prior $S$ contributes a moderate $+0.011$. The per-cell hazard estimate $F_t$ itself costs only $+0.002$.

\textbf{Result on structured landscape.} To test whether this pattern is a property of the default landscape generator (which has no deterministic spatial structure beyond the per-episode wind) or a property of the principle itself, we repeat the ablation on a structured-synthetic landscape: deterministic alternating fuel belts (GR/TL, $4\times$ ROS contrast, 3-row stripes), a diagonal ridge corridor, and an along-belt wind. The FAIR uniform/activity-paced baselines on this landscape are $L_u^{str}=0.075$ and $L_o^{str}=0.042$ (relative gap $44\%$, sanity-gate-passing); the MLP base reaches $L_H^{str}=0.049$ (78\,\% gap closed). The dominant operative component \emph{flips}:

\begin{table}
\caption{\textbf{State-feature ablation on the MLP at $P=2$ on default vs structured landscapes (1200 episodes, 40 test seeds). The dominant operative component flips from temporal staleness (default) to the static-risk prior $S$ (structured); the per-cell hazard estimate $F_t$ remains redundant in both.}}
\label{tab:state}
\setlength{\tabcolsep}{4pt}
\centering
\footnotesize
\begin{tabular}{|l|rr|rr|}
\hline
 & \multicolumn{2}{c|}{Default landscape} & \multicolumn{2}{c|}{Structured landscape} \\
Ablation & $\Delta L_H$ & gap & $\Delta L_H$ & gap \\
\hline
all features (base) & $+0.000$ & $+70$\,\% & $+0.000$ & $+78$\,\% \\
$-$ obs/tx age & $\mathbf{+0.023}$ & $+31$\,\% & $-0.002$ & $+85$\,\% \\
$-$ in-window flag & $\mathbf{+0.024}$ & $+29$\,\% & $-0.002$ & $+85$\,\% \\
$-$ sense-type history & $+0.023$ & $+31$\,\% & $+0.001$ & $+75$\,\% \\
$-$ static-risk prior $S$ & $+0.011$ & $+52$\,\% & $\mathbf{+0.004}$ & $+67$\,\% \\
$-$ per-cell belief $F_t$ & $+0.002$ & $+67$\,\% & $+0.002$ & $+72$\,\% \\
\hline
\end{tabular}
\end{table}

\textbf{Reading.} On the default landscape, the spatial structure is weak (smoothed fuel, random wind), and the scheduler's information lives in when each region was last refreshed: the temporal-staleness axes dominate. On the structured landscape, the spatial structure is strong (deterministic fuel belts and ridges captured by $S$), and the scheduler's information lives in where fires can propagate: the static-risk prior dominates and temporal staleness becomes uninformative. The per-cell intensity belief $F_t$ stays comparatively redundant in both regimes; given the other structured components, the receiver's intensity field is not the primary signal that drives the scheduler. The components of the structured belief track problem structure, consistent with the derivation in Section~\ref{sec:belief}.

\subsection{Realisability: A Modest Architectural Inductive Bias Suffices}
\label{sec:eval-arch}

We test the second half of the principle (maintaining the belief through an anticipatory scheduler) by training the four architectures (MLP, GRU, lightweight cross-region attention, deeper Transformer) under identical online PPO ($P = 2$, $C = 1.0$, $H = 3$, $K = 25$, $N_{\mathrm{BC}} = 1000$, $N_{\mathrm{PPO}} = 4000$).

\begin{table}
\caption{\textbf{Realisability check. Architecture battle on the default landscape, FAIR baseline ($P=2$, $C=1.0$, $H=3$, $K=25$; 5000 PPO episodes including 1000 BC warm-start). FAIR baselines select over the full per-region action space ($s\in\{\text{skip,coarse,fine}\}$, $z\in\{\text{event,summary,frontier,raw}\}$, $t\in\{\text{drop,send}\}$), in contrast with the fixed d* arm of Table~\ref{tab:tempcoup}. All four architectures use $N=5$ training seeds ($s \in \{2031, 2032, 2033, 2034, 2035\}$). $L_H$ entries are mean across training seeds with standard error across training seeds; each per-seed value is itself a mean across 40 evaluation seeds (paired seed set $100$--$139$), taken as the lower of deterministic and stochastic policy evaluation. Gap closure $>100\,\%$ is expected (see Section~\ref{sec:baselines}).}}
\label{tab:arch}
\setlength{\tabcolsep}{4pt}
\centering
\footnotesize
\begin{tabular}{|l|r|c|r|r|}
\hline
Architecture & params & $N$ & $L_H$ & gap closed\\
\hline
FAIR uniform & --- & --- & 0.1505 & 0\,\%\\
FAIR-AP ref.\ & --- & --- & 0.0927 & 100\,\%\\
\hline
MLP & 167\,k & 5 & 0.0908\,$\pm$\,0.0030 & +103.4\,\%\\
GRU & 90\,k & 5 & 0.1057\,$\pm$\,0.0089 & +77.6\,\%\\
\textbf{Lightw.\,Attn.} & \textbf{40\,k} & \textbf{5} & \textbf{0.0670\,$\pm$\,0.0031} & \textbf{+144.5\,\%}\\
Transformer & 202\,k & 5 & 0.0735\,$\pm$\,0.0055 & +133.3\,\%\\
\hline
\end{tabular}
\end{table}

\begin{figure}[!t]
\centering
\includegraphics[width=\columnwidth]{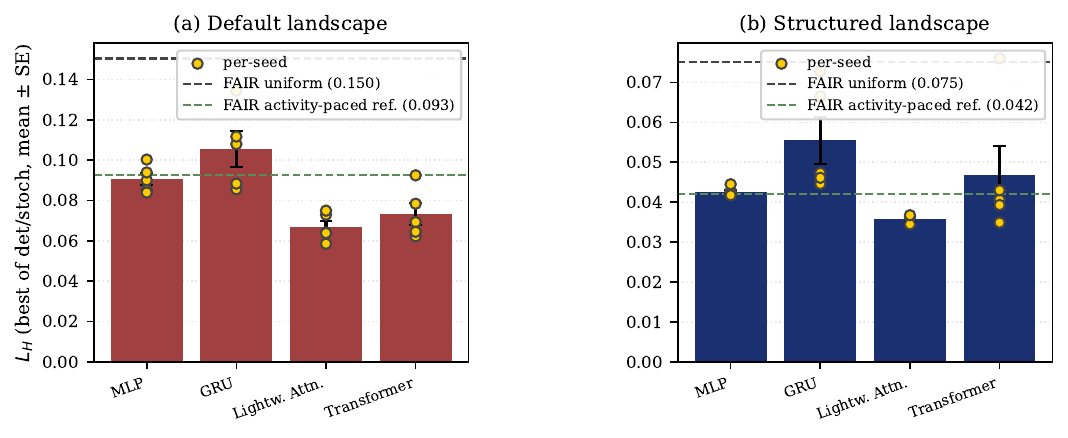}
\caption{\textbf{Realisability check across landscapes. Both panels use matched $N=5$ training seeds for all four architectures; bars show mean $\pm$ SE across training seeds and yellow markers show per-seed values. (a) Default landscape: Lightw.\,Attn.\ ($L_H = 0.0670 \pm 0.0031$\,SE) reaches the lowest mean and exceeds the FAIR activity-paced reference (Section~\ref{sec:baselines}); Transformer ($L_H = 0.0735 \pm 0.0055$\,SE) is close in mean but has higher training-seed variance with one outlier seed at $L_H = 0.0925$; GRU shows the largest training-seed SE (0.0089). (b) Structured landscape: Lightw.\,Attn.\ has the lowest mean; MLP has the smallest training-seed SE (0.0005) and essentially matches the reference; Transformer has the largest training-seed SE (0.0074, one outlier at $L_H = 0.0760$); GRU does not catch up.}\label{fig:arch_struct}}
\end{figure}

\textbf{Result on default landscape (Table~\ref{tab:arch}, Fig.~\ref{fig:arch_struct}(a)).} All four architectures are run under matched seed sets ($N=5$, $s \in \{2031, 2032, 2033, 2034, 2035\}$). The lightweight cross-region attention (40\,k parameters, 2 layers) reaches the lowest mean $L_H = 0.0670 \pm 0.0031$\,SE across training seeds (per-seed $0.0586, 0.0642, 0.0639, 0.0730, 0.0750$), closing $+144.5\,\%$ of the FAIR activity-paced reference gap and is the smallest architecture in our comparison. The deeper Transformer (4 layers, 202\,k parameters) reaches $L_H = 0.0735 \pm 0.0055$\,SE, closing $+133.3\,\%$; four of five Transformer seeds fall in $[0.0623, 0.0785]$ with one outlier at $L_H = 0.0925$ ($s=2033$). The two attention encoders remain close in mean (difference $\approx 0.007$, $\approx 1.0$\,SE combined), with the lightweight encoder lower in mean and tighter in training-seed SE (0.0031 vs 0.0055). The MLP reaches $L_H = 0.0908 \pm 0.0030$\,SE, closing $+103.4\,\%$ and slightly exceeding the activity-paced reference; its training-seed SE is essentially tied with the lightweight encoder (0.0030 vs 0.0031). The GRU reaches $L_H = 0.1057 \pm 0.0089$\,SE, closing $+77.6\,\%$ and exhibits the largest training-seed variance among the four architectures on this landscape (per-seed $0.0856, 0.0884, 0.1080, 0.1118, 0.1345$).

\textbf{Result on structured landscape (Table~\ref{tab:arch_str}, Fig.~\ref{fig:arch_struct}).} We repeat the comparison on the structured landscape (fuel belts, ridge corridor, along-belt wind) defined in Section~\ref{sec:eval-state} under the same $N=5$ seed set.

The lightweight cross-region attention reaches the lowest mean: $L_H^{str} = 0.0374 \pm 0.0012$\,SE over five seeds (per-seed $0.0345, 0.0363, 0.0368, 0.0382, 0.0415$), exceeding the FAIR activity-paced reference in every seed (gap closure $+113.7\,\%$). The MLP is the most stable model under matched seeds: $L_H^{str} = 0.0427 \pm 0.0005$\,SE over five seeds (per-seed $0.0418, 0.0418, 0.0425, 0.0427, 0.0445$; gap closure $+98.0\,\%$), essentially matching the activity-paced reference. Across the two landscapes, the MLP's mean falls just below the FAIR activity-paced reference on the default landscape ($L_H = 0.0908 < L_o^{def} = 0.0927$, $+103.4\,\%$) and just above it on the structured landscape ($L_H^{str} = 0.0427 > L_o^{str} = 0.0420$, $+98.0\,\%$); the lightweight cross-region attention is the only architecture in this comparison whose mean exceeds the FAIR activity-paced reference on both landscapes. The GRU does not catch up: $L_H^{str} = 0.0555 \pm 0.0059$\,SE (per-seed $0.0446, 0.0461, 0.0473, 0.0665, 0.0728$; gap closure $+59.3\,\%$). The deeper Transformer exhibits the largest training-seed variance among the four architectures on this landscape: $L_H^{str} = 0.0468 \pm 0.0074$\,SE, with four of five seeds approaching the lightweight encoder ($L_H \in \{0.0349, 0.0393, 0.0407, 0.0430\}$) and one ($s=2031$) ending at $L_H = 0.0760$, below FAIR uniform; gap closure $+85.6\,\%$. Within the structured landscape and at the 5000-episode budget, the lightweight encoder has the lowest mean $L_H$ and the MLP has the smallest training-seed SE.

\begin{table}
\caption{\textbf{Architecture battle on the structured landscape ($P=2$, $C=1.0$, $H=3$). FAIR baselines select over the full per-region action space (as in Table~\ref{tab:arch}); structured-landscape baselines are $L_u^{str}=0.0751$ and $L_o^{str}=0.0420$. All four architectures use $N=5$ training seeds ($s \in \{2031, 2032, 2033, 2034, 2035\}$). $L_H$ entries are mean across training seeds with standard error across training seeds; each per-seed value is itself a mean across 40 evaluation seeds (paired seed set $100$--$139$). Gap closure $>100\,\%$ is expected (see Section~\ref{sec:baselines}).}}
\label{tab:arch_str}
\setlength{\tabcolsep}{4pt}
\centering
\footnotesize
\begin{tabular}{|l|r|c|r|r|}
\hline
Architecture & params & $N$ & $L_H$ & gap closed\\
\hline
FAIR uniform & --- & --- & 0.0751 & 0\,\%\\
FAIR-AP ref.\ & --- & --- & 0.0420 & 100\,\%\\
\hline
MLP & 167\,k & 5 & 0.0427\,$\pm$\,0.0005 & +98.0\,\%\\
GRU & 90\,k & 5 & 0.0555\,$\pm$\,0.0059 & +59.3\,\%\\
\textbf{Lightw.\,Attn.} & \textbf{40\,k} & \textbf{5} & \textbf{0.0374\,$\pm$\,0.0012} & \textbf{+113.7\,\%}\\
Transformer & 202\,k & 5 & 0.0468\,$\pm$\,0.0074 & +85.6\,\%\\
\hline
\end{tabular}
\end{table}

\textbf{Reading.} Read as a realisability and stability check under matched $N=5$ seed sets, the architecture comparison supports four observations. First, MLP is the most stable architecture under matched seeds (lowest training-seed SE on both landscapes: 0.0030 default, 0.0005 structured), and on both landscapes its mean is close to the FAIR activity-paced reference; a memoryless state-rich policy thus realises the principle when the belief is correctly posed. Second, GRU does not catch up to the activity-paced reference on either landscape (gap closure $+77.6\,\%$ default, $+59.3\,\%$ structured) and exhibits the largest training-seed variance among the four architectures on the default landscape (per-seed range $0.0856$--$0.1345$); recurrent memory in our setup is not a free improvement, and the single-seed appearance in earlier reporting understated this training-seed variance. Third, the lightweight cross-region attention has the lowest mean $L_H$ on both landscapes and is the smallest architecture (40\,k parameters); its training-seed SE is essentially tied with MLP on default (0.0031 vs 0.0030) and is higher than MLP's on structured (0.0012 vs 0.0005). Fourth, the deeper Transformer is high-variance on both landscapes: on default, four of five seeds approach the lightweight encoder (mean difference $\approx 0.007$, $\approx 1.0$\,SE combined) but one seed lands at $L_H = 0.0925$; on structured it has the largest training-seed SE of the four architectures (0.0074), with four of five seeds approaching the encoder and one seed failing entirely at $L_H = 0.0760$. The defensible reading is therefore narrower than "attention is necessary": under this task class and this regime, when the design principle of Section~\ref{sec:principle} is realised in the right regime (Section~\ref{sec:eval-tempcoup}) with the right structured belief (Section~\ref{sec:eval-state}), a modest architectural inductive bias suffices to realise the principle, and adding depth is neither necessary nor automatically helpful. Gap closure above 100\,\% is expected (see Section~\ref{sec:baselines}). Architecture comparison in this paper is a realisability and stability check, not a statement about which inductive bias is essential.

\begin{figure}[!t]
\centering
\includegraphics[width=\columnwidth]{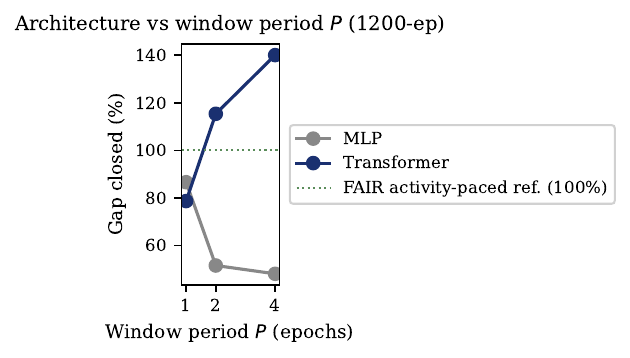}
\caption{\textbf{Architecture--sparsity interaction at the 1200-ep budget, with MLP and Transformer as architectural endpoints. The Transformer's gap-closure grows with $P$ while the MLP plateaus near $+50$\,\%; the architectural-bias advantage grows with the temporal-coupling regime, consistent with the principle that opportunity-aware scheduling has its empirical room at intermediate sparsity. Gap closure $>100\,\%$ is expected (see Section~\ref{sec:baselines}). The lightweight encoder's 5000-ep value at $P=2$ is in Table~\ref{tab:arch}; see Table~\ref{tab:budget} for the budget dependence of all architectures.}\label{fig:p_arch}}
\end{figure}

\textbf{Sparsity interaction.} Fig.~\ref{fig:p_arch} reports a secondary observation at the 1200-episode budget. Here we use the deeper Transformer and the MLP as architectural endpoints (sample-efficient bias vs.\ memoryless input), with the lightweight encoder's headline value already in Table~\ref{tab:arch} at $P=2$. The gap-closure advantage of the Transformer over the MLP grows monotonically with window period $P$: at $P = 1$ the MLP matches; at $P = 2$ and $P = 4$ the Transformer's lead reaches $+115$\,\% and $+140$\,\% gap-closure while the MLP plateaus near $+50$\,\% (gap closure $>100\,\%$ as expected; see Section~\ref{sec:baselines}). The architectural advantage matters precisely in the temporal-coupling regime where opportunity-aware scheduling has empirical room.

\subsection{Honest Comparison: FAIR vs.~UNFAIR Baselines}
\label{sec:eval-fair}

Table~\ref{tab:fair} compares the UNFAIR (privileged-truth) and FAIR (belief-only) baseline schedulers at the operating point. The UNFAIR uniform baseline reaches $L_H = 0.059$ by scoring regions against the true fire field. The FAIR uniform baseline reaches $L_H = 0.150$ using only the deployable belief, a $2.5\times$ inflation under the same scheduling algorithm and pacing. A learned policy that closes a substantial fraction of the UNFAIR activity-paced gap may be reproducing baseline information leakage rather than learning improvement. Learned-policy results in this paper are reported against the FAIR (deployable) baseline. The rank ordering in $L_H$ is preserved in the secondary binary metrics (F1, IoU), with the relative gap smaller in absolute terms because the binary metric saturates near 0.85--0.90 once the fire field has unambiguously burning and non-burning cells.

\begin{table}
\caption{\textbf{Honest comparison: UNFAIR vs.~FAIR baselines at $P=2$, $C=1.0$, $H=3$ ($K=25$; 40 seeds). UNFAIR baselines have privileged access to the true fire field; FAIR baselines use only the deployable belief. Both UNFAIR and FAIR baselines select over the full per-region action space; cf. the fixed-action d* arm of Table~\ref{tab:tempcoup}. For binary thresholded classification the Dice coefficient is mathematically identical to the F1 score, so we report only F1 and IoU.}}
\label{tab:fair}
\setlength{\tabcolsep}{6pt}
\centering
\footnotesize
\begin{tabular}{|l|r|r|r|}
\hline
 & $L_H$ & F1 & IoU \\
\hline
UNFAIR uniform & 0.059\,$\pm$\,0.004 & 0.904 & 0.857 \\
UNFAIR activity-paced & 0.062\,$\pm$\,0.004 & 0.900 & 0.853 \\
UNFAIR eager & 0.073\,$\pm$\,0.006 & 0.877 & 0.817 \\
\hline
FAIR uniform & 0.150\,$\pm$\,0.013 & 0.856 & 0.811 \\
\textbf{FAIR-AP ref.\ } & \textbf{0.093}\,$\pm$\,0.009 & \textbf{0.885} & \textbf{0.836} \\
FAIR eager & 0.147\,$\pm$\,0.012 & 0.857 & 0.812 \\
\hline
\end{tabular}
\end{table}

\subsection{Budget Dependence of the Architecture Ranking}
\label{sec:eval-budget}

\begin{figure}[!t]
\centering
\includegraphics[width=\columnwidth]{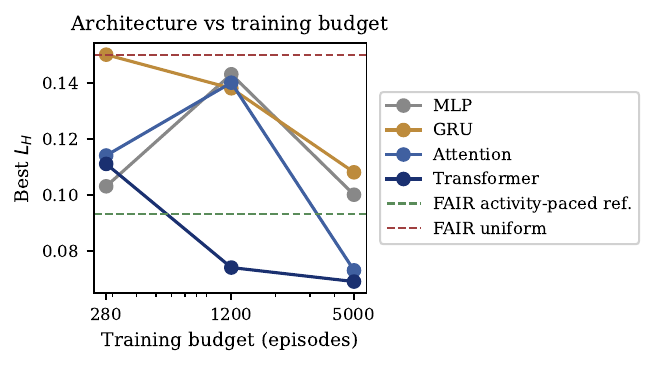}
\caption{\textbf{Budget-dependent realisation. At small budgets (280\,ep) the simpler MLP is competitive; at moderate budgets (1200\,ep) the deeper Transformer is realised first; at sufficient budgets (5000\,ep) the lightweight attention encoder realises the principle at modest cost.}\label{fig:budget_curve}}
\end{figure}

The realisability claim depends on the training budget. Table~\ref{tab:budget} and Fig.~\ref{fig:budget_curve} compare the best $L_H$ at three budgets.

\begin{table}
\caption{\textbf{Budget dependence of the realisation. Best $L_H$ across deterministic/stochastic; lower is better.}}
\label{tab:budget}
\setlength{\tabcolsep}{6pt}
\centering
\footnotesize
\begin{tabular}{|l|r|r|r|}
\hline
Architecture & 280 ep & 1200 ep & 5000 ep\\
\hline
MLP & 0.103 & 0.143 & 0.100\\
GRU & 0.150 & 0.138 & 0.108\\
Lightw.\,Attn. & 0.114 & 0.140 & \textbf{0.073}\\
Transformer & 0.111 & \textbf{0.074} & \textbf{0.069}\\
\hline
\end{tabular}
\end{table}

\textbf{Reading.} At small budgets the simpler MLP is competitive; at the moderate budget only the deeper Transformer realises the principle (the lightweight attention has not yet recovered from the under-trained regime); at the full budget the lightweight attention realises the principle at modest cost. The implication is that realisability is a function of training budget as well as architecture, and that any realisability claim about belief-aware scheduling should be reported with explicit budget annotation. This does not contradict the main claim: the deeper Transformer needs more samples to realise the principle, while the lightweight encoder realises it at fewer samples but still wants enough; the budget-dependent ranking is a sample-efficiency effect under online PPO, and the headline result is the 5000-episode behaviour where the lightweight encoder is both reliable and sufficient (Section~\ref{sec:eval-arch}). Gap closure $>100\,\%$ read off Table~\ref{tab:budget} via the percentage definitions of Table~\ref{tab:percentdef} is expected (see Section~\ref{sec:baselines}).

\subsection{Robustness across Predictive Horizons}
\label{sec:eval-horizon}

\begin{figure}[!t]
\centering
\includegraphics[width=\columnwidth]{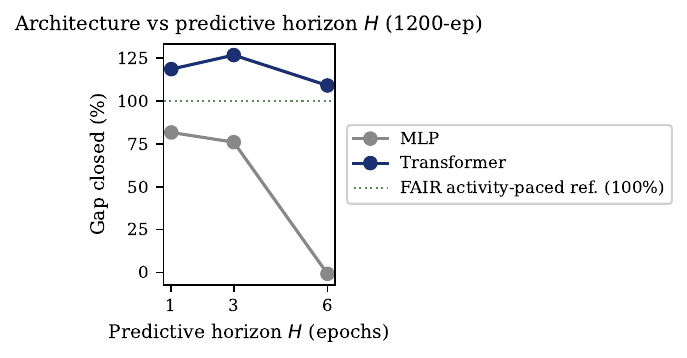}
\caption{\textbf{H-axis ablation at the 1200-ep budget, with MLP and Transformer as architectural endpoints. The Transformer's gap closure remains above the FAIR activity-paced reference (green dashed) across all $H \in \{1, 3, 6\}$; the MLP plateaus and at $H = 6$ is effectively tied with this reference. Gap closure $>100\,\%$ is expected (see Section~\ref{sec:baselines}). The lightweight encoder's 5000-ep value at $H = 3$ is in Table~\ref{tab:arch}; see Table~\ref{tab:budget} for the budget dependence.}\label{fig:h_arch}}
\end{figure}

The principle should be robust across predictive horizons $H$. Fig.~\ref{fig:h_arch} reports the H-ablation, again using the Transformer and MLP as architectural endpoints at the 1200-episode budget (the lightweight encoder's value at $H=3$ is in Table~\ref{tab:arch}). The Transformer beats the FAIR activity-paced reference across all three horizons ($+119$\,\%, $+127$\,\%, $+109$\,\% gap closure at $H = 1, 3, 6$; gap closure $>100\,\%$ as expected, see Section~\ref{sec:baselines}). The MLP closes $+82$\,\%, $+76$\,\%, and $-0.8$\,\% respectively; at $H = 6$ the FAIR uniform $L_H$ drops because long-horizon predictions naturally smooth, so the MLP plateaus near the FAIR activity-paced reference while the Transformer maintains a clear lead. The realisation is robust to $H$ at the 1200-episode budget.

\section{Discussion and Limitations}
\label{sec:disc}

\textbf{What we do not claim.} Cross-region attention is not claimed to be the necessary inductive bias for this class of problem. The architectural comparison is a realisability check, not a statement that spatial reasoning is essential.

\textbf{Structure-dependence of the operative belief.} The reading that ``the dominant scheduling levers are temporal'' on the default landscape might be an artefact of the absence of deterministic spatial structure. Section~\ref{sec:eval-state} tests this directly. On a structured-synthetic landscape (fuel belts, ridge corridor, along-belt wind) the dominant operative component shifts from temporal staleness ($+0.023$ on default) to the static-risk prior $S$ ($+0.004$ on structured), while the per-cell intensity belief $F_t$ remains comparatively redundant in both. The components of $B_t$ track problem structure rather than being fixed, and the principle is unchanged.

\textbf{Real-landscape evaluation.} Real-landscape data introduces jointly-varying fuel, terrain, wind, and ignition statistics. Evaluating belief-aware scheduling under those joint confounds is a distinct study from the one reported here, whose design (Section~\ref{sec:scenario}) requires independent control of $P$, $C$, $H$, and fuel composition. Integration of LANDFIRE / ERA5 / NIFC ingestion is a natural future direction. The structured-synthetic landscape used in Section~\ref{sec:eval-state} isolates the role of deterministic spatial structure within the present design.

\textbf{Sparse-window model.} The periodic-window model is the tier-(iv) controlled abstraction in Section~\ref{sec:runtime}'s hierarchy. A natural follow-up is to rerun the architecture battle under stochastic availability $A_t \in \{0,1\}$ with the expected number of windows matched to the periodic setting (tier (iii)). The structured-landscape test already exercises the principle's most stressful axis, so this stochastic-availability check is a lower-priority follow-up.

\textbf{Belief representation.} Each component of $B_t$ is a point estimate. Bayesian belief filtering (particle filter over $F_t$, covariance over $U_t$, neural belief from temporal attention) is a natural extension that may improve uncertainty-aware scheduling under high observation noise or sparse sensing. Such extensions refine the realisation, not the principle.

\textbf{Forward operator assumed known.} The principle treats $\Phi(\cdot, H)$ as a fixed, known forward operator (here, the Rothermel-calibrated dynamics integrated for $H$ epochs). Model mismatch and forward-operator misspecification are out of scope: a learned or partially-learned $\Phi$ would change the derivation rule for $B_t$ (uncertainty over $\Phi$ itself would enter the belief), but our claims do not address this. Studying belief-aware scheduling under uncertain $\Phi$ is future work.

\textbf{Training budget.} The learned-policy experiments use up to 5000 PPO episodes ($\approx$125,000 environment steps), well below the 5M+ steps standard in POMDP RL literature~\cite{morad2023popgym}. The realisability claim holds at the 1200- and 5000-episode budgets and does not hold at a 280-episode pilot budget where the MLP dominates. Extension to SOTA training scale with vectorised environments is left as future work.

\textbf{Scope.} The single-drone scope isolates the temporal-opportunity coupling. Multi-drone swarm coordination with per-agent constraints and inter-drone communication is a distinct research question with additional spatial coupling not characterised here.

\section{Conclusion}
\label{sec:conc}
We articulated belief-aware scheduling as a derivation rule for predictive telemetry under sparse-window constraints, and, for the task class and regime studied (single-drone predictive wildfire hazard mapping with periodic windowed transmission and a pooled budget), supported it with three empirical findings: a unimodal activity-paced--uniform gap in window-period sparsity (the regime where the principle has empirical room); a per-component belief ablation whose dominant component flips between landscapes (evidence that the decomposition tracks problem structure rather than being engineered); and a realisability check in which a 40\,k-parameter lightweight cross-region attention encoder suffices on both landscapes, while a deeper Transformer offers no reliable advantage. We therefore find that within this task class and regime, when the belief and the scheduling problem are correctly posed (per the derivation rule), a modest architectural inductive bias is sufficient; we do not claim that this conclusion transfers without re-validation to other predictive tasks, other connectivity regimes (e.g., stochastic or non-periodic availability), or larger architectures with different training budgets.

\section*{Acknowledgment}
This work was partly supported by JSPS KAKENHI 24K14913, the Telecommunications Advancement Foundation (SCAT), and the Joint Research Project of the Research Institute of Electrical Communication, Tohoku University.

\bibliographystyle{IEEEtran}
\bibliography{references}

\appendices
\section{Architecture Implementation Details}
\label{app:arch}
Per-architecture hyperparameters: MLP (167\,k params, two hidden layers of 200 units, ReLU); GRU (90\,k params, 80-unit GRU cell, single-layer head); Lightweight Attention (40\,k params, 2 transformer encoder layers, embedding dim $d = 48$, 4 heads, feedforward dim 96, learned positional embeddings over the 25-region sequence); Transformer (202\,k params, 4 layers, $d = 64$, 4 heads, feedforward dim 256). All architectures use per-region categorical action heads with three softmax outputs (sensing, representation, transmission) and share the structured-belief input.

\section{Deterministic and Stochastic Evaluation Values}
\label{app:detstoch}

The $L_H$ values reported in Tables~\ref{tab:arch}--\ref{tab:arch_str} use the \textit{best of det/stoch} convention (Section~\ref{sec:eval-settings}): for each training seed, the lower of the deterministic and stochastic policy-evaluation means (across 40 evaluation seeds) is taken as that seed's $L_H$, and the cross-seed mean of these per-seed best values is reported. Table~\ref{tab:detstoch} shows both the per-seed det and per-seed stoch means side-by-side, alongside the resulting best-of-det/stoch value used in the main tables.

\begin{table}
\caption{\textbf{Deterministic and stochastic evaluation means per architecture per landscape}, both reported as mean $\pm$ SE across training seeds (each per-seed entry is itself a mean across 40 evaluation seeds). The ``best'' column is the per-seed minimum of det and stoch, then averaged across seeds, and matches the values reported in Tables~\ref{tab:arch}--\ref{tab:arch_str}.}
\label{tab:detstoch}
\setlength{\tabcolsep}{4pt}
\centering
\footnotesize
\begin{tabular}{|l|l|c|r|r|r|}
\hline
Landscape & Architecture & $N$ & $L_H$ (det) & $L_H$ (stoch) & best\\
\hline
Default & MLP & 5 & 0.1366\,$\pm$\,0.0137 & 0.0908\,$\pm$\,0.0030 & 0.0908\\
Default & GRU & 5 & 0.1084\,$\pm$\,0.0097 & 0.1122\,$\pm$\,0.0064 & 0.1057\\
Default & Lightw.\,Attn. & 5 & 0.0825\,$\pm$\,0.0137 & 0.0701\,$\pm$\,0.0019 & 0.0670\\
Default & Transformer & 5 & 0.1192\,$\pm$\,0.0398 & 0.0742\,$\pm$\,0.0053 & 0.0735\\
\hline
Structured & MLP & 5 & 0.0462\,$\pm$\,0.0040 & 0.0450\,$\pm$\,0.0003 & 0.0427\\
Structured & GRU & 5 & 0.0555\,$\pm$\,0.0059 & 0.0586\,$\pm$\,0.0061 & 0.0555\\
Structured & Lightw.\,Attn. & 5 & 0.0374\,$\pm$\,0.0012 & 0.0392\,$\pm$\,0.0011 & 0.0374\\
Structured & Transformer & 5 & 0.0570\,$\pm$\,0.0118 & 0.0485\,$\pm$\,0.0073 & 0.0468\\
\hline
\end{tabular}
\end{table}

In every row of Table~\ref{tab:detstoch}, det and stoch column means are within $1$--$2$\,SE of each other (excluding the MLP and Transformer Default-landscape det means, where seeds with sharply-collapsed deterministic distributions inflate the det average); the best-of-det/stoch convention absorbs this incidental difference rather than selecting between competing measures of policy quality.

\begin{IEEEbiography}[{\includegraphics[width=1in,height=1.25in,clip,keepaspectratio]{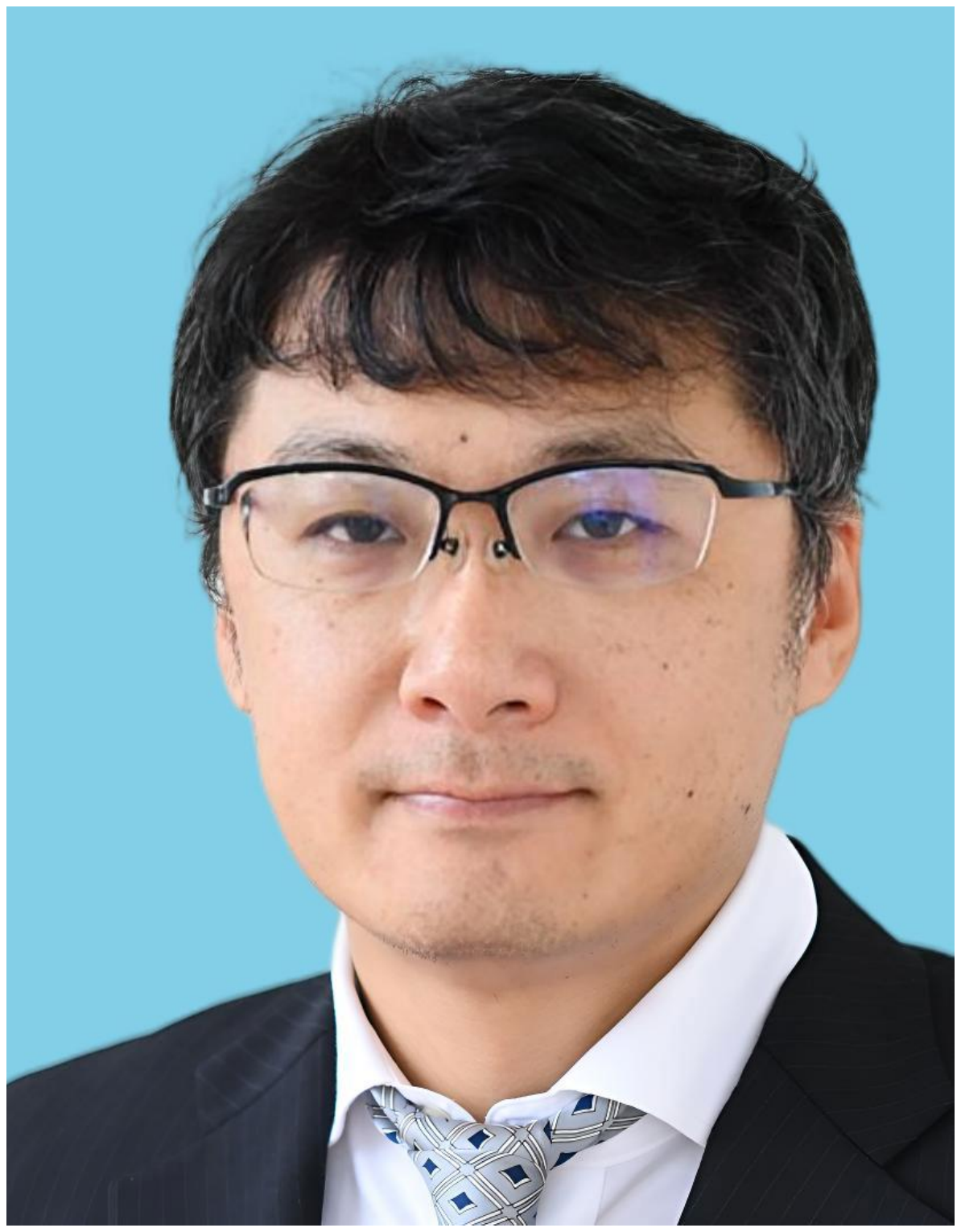}}]{Xun Shao}
(Senior Member, IEEE) was a researcher with the National Institute
of Information and Communications Technology (NICT), Japan. From
2018 to 2022, he was an Assistant Professor with the Kitami
Institute of Technology, Japan. He is currently an Associate
Professor with the Department of Electrical and Electronic
Information Engineering, Toyohashi University of Technology,
Toyohashi, Japan. His research interests include distributed
systems and information networking
(e-mail: shao.xun.ls@tut.jp).
\end{IEEEbiography}

\begin{IEEEbiography}[{\includegraphics[width=1in,height=1.25in,clip,keepaspectratio]{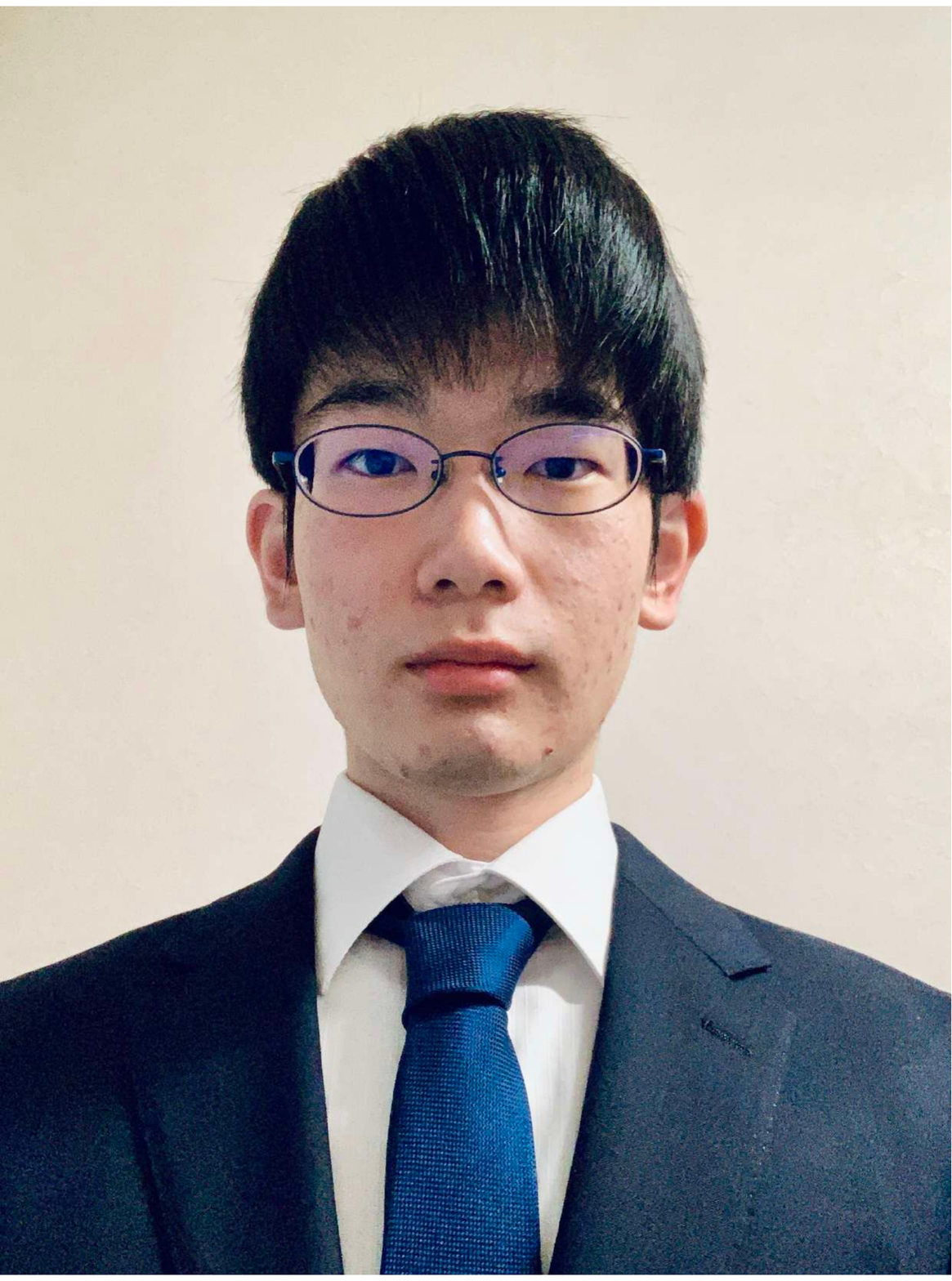}}]{Kohsuke Yamakawa}
is a master student in the Department of
Electrical and Electronic Information Engineering, Toyohashi
University of Technology. His current research interests include
deep learning, computer networks, and cloud computing
(e-mail: yamakawa.kosuke.wy@tut.jp).
\end{IEEEbiography}

\begin{IEEEbiography}[{\includegraphics[width=1in,height=1.25in,clip,keepaspectratio]{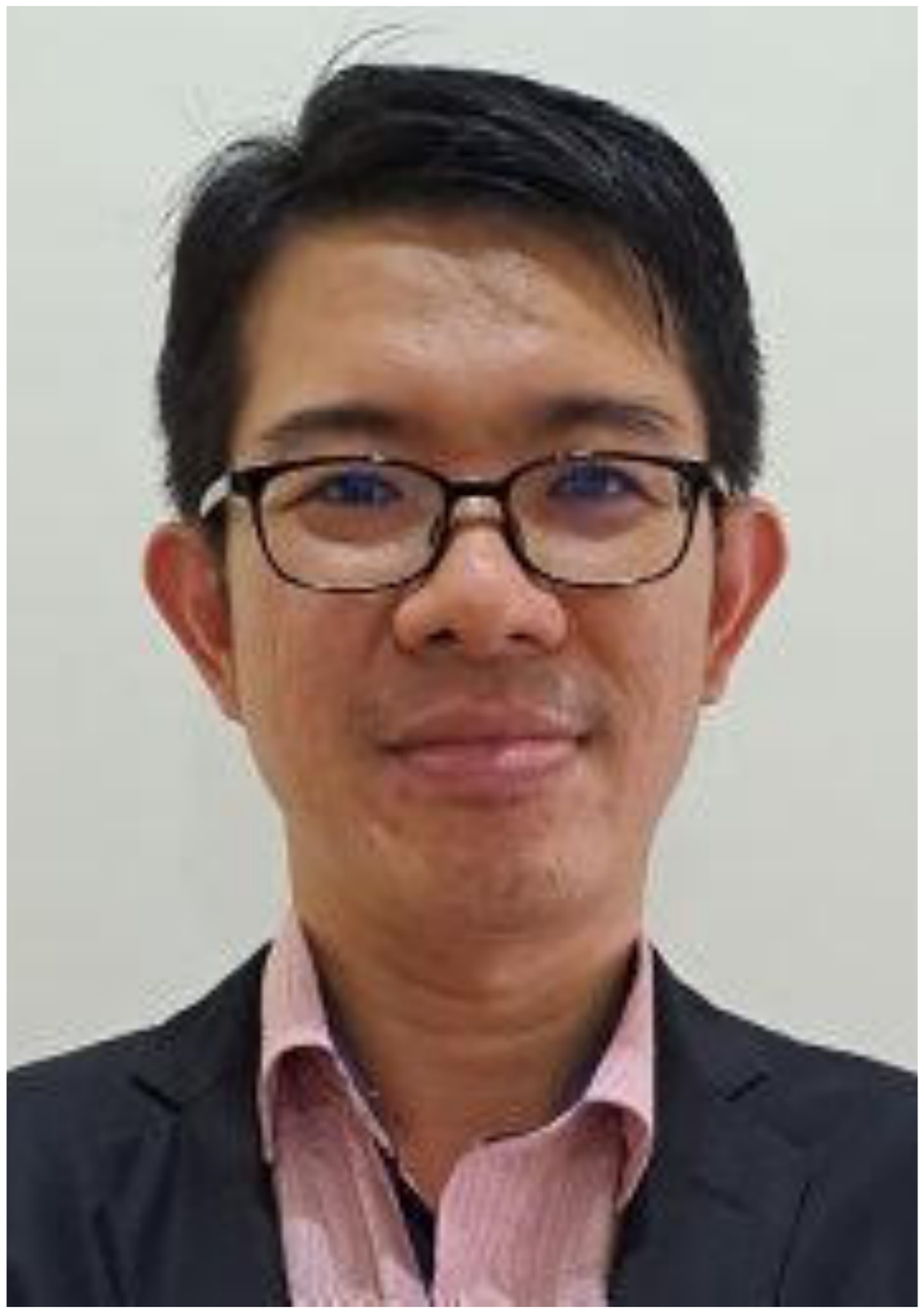}}]{Cheah Wai Shiang}
is an Associate Professor at the Faculty of Computer Science and Information Technology, Universiti Malaysia Sarawak (UNIMAS). He received the Master of Network Computing from Monash University, Australia, and the Ph.D. in Software Engineering from The University of Melbourne, Australia, in 2010. From 2014 to 2016, he pursued postdoctoral research at Utrecht University and VU Amsterdam, the Netherlands. He is a certified Android developer. His main research interests include agents in real-time environments, particularly agent technology and games AI, and validating Agent-Oriented Modelling (AOM). He is currently working on edge AI and exploring quantum computing.
\end{IEEEbiography}

\end{document}